\documentclass[12pt]{article}

\usepackage{graphicx}
\usepackage{graphics}
\usepackage{epsfig}
\usepackage{amssymb}
\usepackage{amsmath}
\usepackage{color}
\usepackage{textcomp}
\usepackage{cite}

\usepackage[margin=2.5cm,footskip=0.6cm]{geometry}
\begin{document}

\begin{center}
{\Large{\bf Universality of Domain Growth in Antiferromagnets with Spin-Exchange Kinetics}} \\
\ \\
\ \\
by \\
Prasenjit Das$^1$, Tanusri Saha-Dasgupta$^{2}$ and Sanjay Puri$^1$ \\
$^1$ School of Physical Sciences, Jawaharlal Nehru University, New Delhi 110067, India. \\
$^2$ S. N. Bose National Centre for Basic Sciences, Kolkata 700098, India. \\
\end{center}

\begin{abstract}
\noindent We study phase ordering kinetics in symmetric and asymmetric binary mixtures, undergoing an order-disorder transition below the critical temperature. Microscopically, we model the kinetics via antiferromagnetic Ising model with Kawasaki spin-exchange kinetics. This conserves the composition while the order-parameter (staggered magnetization) is not conserved. The order-parameter correlation function and structure factor show dynamical scaling, and the scaling functions are independent of the mixture composition. The average domain size shows a power-law growth: $L_\sigma(t)\sim t^\alpha$. The asymptotic growth regime has $\alpha=1/2$, though there can be prolonged transients with $\alpha<1/2$ for asymmetric mixtures. Our unambiguous observation of the asymptotic universal regime is facilitated by using an accelerated Monte Carlo technique. We also obtain the coarse-grained free energy from the Hamiltonian, as a function of two order-parameters. The evolution of these order-parameters is modeled by using \textit{Model C} kinetics. Similar to the microscopic dynamics, the average domain size of the nonconserved order-parameter (staggered magnetization) field exhibits a power-law growth: $L_m(t)\sim t^{1/2}$ at later times, irrespective of the mean value of the conserved order-parameter (composition) field.
\end{abstract}

\newpage
\section{Introduction}\label{3sec1}
\textit{Phase ordering} processes are of great interest in the fields of materials science and metallurgy for the designing of new materials~\cite{pw2009,sdsp04}. Apart from this, they present a fascinating class of problems in thermodynamics and phase transitions~\cite{pw2009,ajb94,mpbb06}. When a disordered binary mixture A$_x$B$_{1-x}$ is suddenly quenched below the order-disorder critical temperature, $T_c(x)$, the system evolves toward two degenerate (ABAB and BABA) ordered domains~\cite{kdm03,fsg08}. The appropriate order parameter to describe this transition is the {\it staggered magnetization}. The mixture is said to be symmetric if $x=0.50$, otherwise it is asymmetric. The kinetics of order-disorder transitions in symmetric mixtures is well-studied by using both microscopic and coarse-grained models~\cite{kygs83,gss84}. At the microscopic level, the order-disorder transition is investigated by using the nearest-neighbor~(nn) Ising antiferromagnet with Kawasaki spin-exchange kinetics. The average domain size $L(t)$ grows with time as $L(t)\sim t^{1/2}$, corresponding to diffusive growth with nonconserved order parameter (Allen-Cahn theory)~\cite{ca1979}. The domain morphology is studied by using the order-parameter correlation function $C(r, t)$, and it's Fourier transform, the structure factor $S(k, t)$. These quantities show dynamical scaling with the scaling forms~\cite{pw2009}: 
\begin{align}
 \label{eqn0}
 C(r, t) &= g\left(r/L\right),\\
 S(k, t)&= L^df(kL),
\end{align}
where $d$ is the spatial dimensionality. Here, $g(x)$ and $f(p)$ are master functions, which are independent of time. For the symmetric mixture, the scaling function $g(x)$ is well-defined by the {\it Ohta-Jasnow-Kawasaki}~(OJK) function~\cite{ojk82} for nonconserved kinetics.

The kinetics of ordering has been studied experimentally for both symmetric and asymmetric compositions~\cite{nshs88,bbl90,snhn92}. The first experimental study of ordering in a symmetric Cu$_3$Au mixture, which has FCC structure, is due to Hashimoto~\textit{et al.}~\cite{hnt78}. They found that the average domain size grows very slowly in the early stages, with a crossover to the $t^{1/2}$-law at later times. Later, Katano~\textit{et al.}~\cite{kinc88} investigated the kinetics of atomic ordering in Ni$_3$Mn FCC alloy by using time-resolved neutron-diffraction techniques. In their study, the length scale of ordered domains shows a crossover from a $t^{1/4}$-law at early times to a $t^{1/2}$-law at later times. Further, the structure factor follows dynamical scaling. Malis and Ludwig~\cite{ml1999} have studied the ordering kinetics in a symmetric CuAu mixture, which has BCC structure. They reported a $t^{1/2}$-growth-law in the late stages of coarsening. 

The first experimental study of ordering kinetics in asymmetric mixtures is due to Shannon~\textit{et al.}~\cite{shn88}. They studied ordering kinetics in sputtered films of Cu$_{0.75+x}$Au$_{0.25-x}$. Below $T_c$, for symmetric compositions with $x=0$, domain coarsening is consistent with a $t^{1/2}$-law. The domain growth in asymmetric films with $x=0.04$ is much slower and shows a logarithmic time dependence. The ``logarithmic'' regime could be a consequence of quenched impurities in the system~\cite{ppr0405, lmpz10,clmp13}. It is also possible that it could be a transient regime prior to an asymptotic power-law growth. Rivers~\textit{et al.}~\cite{ruhl95} studied order-disorder transition at the (001) surface of Cu$_{0.78}$Au$_{0.22}$ crystal by using surface x-ray scattering and Auger-electron spectroscopy. They also reported a bulk domain growth slower than $t^{1/2}$.

The ordering kinetics in symmetric and asymmetric binary mixtures has been studied via Monte Carlo (MC) simulations \cite{ohta84, gfl97}. For symmetric mixtures on a simple cubic lattice, Phani and Lebowitz~\cite{pl1980} showed that the characteristic length increases as $t^{1/2}$. Subsequently, Sahni~\textit{et al.}~\cite{sd1981} showed that the structure factor follows dynamical scaling. The first MC study of ordering in asymmetric mixtures is due to Porta and Cast\'{a}n~\cite{pc1996}. They studied the effect of composition asymmetry on ordering in A$_x$B$_{1-x}$ mixtures, with $x\leq0.50$ in $d=2$ square lattices. They found that the characteristic scale follows a power-law: $L(t)\sim t^\alpha$, with exponent $\alpha\sim 0.40-0.50$ for $x\in[0.4,0.5]$. However, they predicted the growth law $L(t)\sim t^{1/2}$, irrespective of the asymmetry. A similar growth law was also observed by Frontera~\textit{et al.}~\cite{fvcp97} for L1$_2$-ordered domains in FCC A$_3$B binary alloys.

There also exist a few coarse-grained model studies of ordering in symmetric and asymmetric mixtures~\cite{vgv04}. Lai~\cite{lai90} constructed a coarse-grained model to describe the ordering kinetics in A$_3$B alloy on FCC lattices. His model is characterized by a \textit{Ginzburg-Landau}~(GL) Hamiltonian with a three-component order parameter and the symmetry of the A$_3$B system. Results obtained from this model are in good agreement with the experimentally observed growth kinetics. Somoza and Sagui~\cite{ss1996} studied a \textit{Model C} system, which is defined by the coupled equations for a nonconserved order parameter (staggered magnetization) and  a conserved variable (composition). They argued that the wetting properties of interfaces between differently ordered domains modify the domain growth morphology. Later, Kockelkoren and Chat\'{e} \cite{kc2002} studied the late stages of coarsening in a \textit{Model C} system. They reported a $t^{1/2}$-domain-growth law in the nonconserved order-parameter field. Subsequently, Gumennyk~\textit{et al.}~\cite{gsf09} obtained the evolution equations for non-conserved and conserved order-parameters for ordering in BCC lattice structures. They started with coupled mean-field kinetic equations for relaxation of occupancies in the two sublattices. They observed that long range order formed at the early stage of evolution, and was followed by the slow redistribution of alloy concentrations. However, they did not study the morphological features of domain growth and the growth law of ordering kinetics.

In real alloys, ordering dynamics is driven by vacancy (V)-mediated exchanges of atoms rather than direct exchanges as considered in the above studies. The first MC study of ordering with this mechanism is due to Yaldram and Binder~\cite{kykb1,kykb2}. Depending upon the strength of mutual interactions among A, B and V, and their concentrations, they observed uniformly distributed vacancies in the system as well as enrichment of vacancies along the domain interfaces. Later, Puri and Sharma~\cite{sprs98} formulated mean-field dynamical models for segregation in binary mixtures driven by vacancies. In their study, the average domain size grows as $L\sim t^{1/3}$. Le Floc'h, Bellon \textit{et al.}~\cite{dld200,jia204} studied the ordering of B2-ordered domains by using MC simulations. They also observed a $t^{1/2}$-growth-law for average domain size. It is believed that vacancy-mediated ordering and the direct-exchange mechanism give the same growth law and differ only by a renormalization of the time-scale. Most recently, the first-principles study of Sanati and Zuger~\cite{sanati} also confirms that late-stage domain growth in Cu$_3$Au FCC alloys mediated by vacancies follows the same law.

In spite of the above mentioned studies, the understanding of ordering kinetics in asymmetric mixtures remains incomplete. Firstly, there is lack of clarity regarding the asymptotic growth regime due to slow transients. Secondly, it is not clear whether the domain morphologies are dependent on the level of asymmetry. Thirdly, which is related to the second issue, it is not known whether the surplus component wets the domain boundaries for asymmetric mixtures, or it dissolves into the bulk domains. In this paper, we attempt to address these issues via MC simulations of the kinetics of order-disorder transitions in A$_x$B$_{1-x}$ mixtures in $d=2$. In particular, we use an accelerated MC algorithm to unambiguously demonstrate the following:\\
1. The asymptotic domain growth law is $L(t)\sim t^{1/2}$, regardless of the composition of the mixture.\\
2. The scaling functions, which characterize the evolution morphologies, are universal for different values of $x$.\\
3. While at the initial stage of the growth, the surplus component does wet the domain boundaries, at the later stage the surplus component migrate into the bulk domains. This in turn reduces the value of the staggered magnetization. \\
Our MC results are complemented by results from a coarse-grained kinetic model ({\it Model~C}). These latter results independently verify the above assertions. We believe that our results in $d=2$ will also be valid for different crystal structures and higher-component order parameters in $d=3$.

This paper is organized as follows. In Sec.~\ref{3sec2}, we give details of the microscopic model and its MC simulations. The model is discussed in Sec.~\ref{3s1sec2}, and detailed numerical results are presented in Sec.~\ref{3s2sec2}. In Sec.~\ref{3sec3}, we present the corresponding coarse-grained model~(Sec.~\ref{3s1sec3}) and results obtained therefrom~(Sec.~\ref{3s2sec3}). Finally, we conclude with a summary and discussion in Sec.~\ref{3sec4}.

\section{Antiferromagnets with Kawasaki Kinetics}\label{3sec2}
\subsection{Model and Numerical Details}\label{3s1sec2}
For the microscopic model, we consider the Ising antiferromagnet:
\begin{eqnarray}
 \label{eqn1}
 \mathcal{H} = J\sum_{\langle ij\rangle}S_iS_j, ~~~~~~~S_i=\pm 1.
\end{eqnarray}
In Eq.~(\ref{eqn1}), $J$ is the strength of the exchange interaction, which is positive (i.e. $J>0$) for the order-disorder transition. The subscript $\langle ij\rangle$ represents nn interactions, and $S_i = +1$ or $-1$ corresponds to A or B atoms of the binary mixture respectively. We associate kinetics with this Ising model by placing the system in contact with a heat bath that generates a stochastic exchange of atoms (A$\leftrightarrow$B) between two neighboring sites. This model is known as the {\it Kawasaki spin-exchange} model \cite{pw2009}. Below $T_c$, the A and B atoms order on alternate sublattices. The appropriate order parameter is the {\it staggered magnetization M}, which is the difference between two sublattice \textit{magnetizations}. It is not conserved under the Kawasaki kinetics, though the system has a conserved quantity, viz., the composition.

We study ordering dynamics for this in $d=2$. We choose the system size ($L^2$) to be $4096\times 4096$. We employ periodic boundary conditions in all the directions. The initial condition of our MC simulation consists of a random distribution of A and B with number densities $c_A$ and $c_B(=1 - c_A)$, mimicking the disordered state before the quench. At $t=0$, the system was quenched to $T<T_c$. A randomly chosen pair of unlike spins are interchanged according to the above stochastic move, corresponding to a change in configuration from $\{S_i\}\rightarrow\{S_i^{'}\}$. The change is accepted with probability $p$, given by
\begin{align}
\label{mcs}
  p = \left\{
  \begin{array}{lr}
    1 & : \Delta E\le0,\\
    \text{exp}(-\beta\Delta E) & : \Delta E>0.
  \end{array}
\right.
\end{align}
Here, $\Delta E = \mathcal{H}(\{S_i^{'}\}) - \mathcal{H}(\{S_i\})$ is the energy difference between the final and initial configurations and $\beta=(k_BT)^{-1}$ with $k_B=1$~\cite{DLKB15}. One Monte Carlo step (MCS) corresponds to $L^2$ attempted updates. All the statistical results presented here are obtained as averages over ten independent runs.

The standard MC approach described above is not very useful in accessing the asymptotic regimes of antiferromagnetic ordering, particularly for asymmetric compositions. For asymmetric mixtures, the excess component initially wets the interfaces, resulting in a drastic slowing down of domain growth. To overcome this problem, we use an accelerated approach introduced by Marko and Barkema~(MB) in the context of phase-separating binary alloys~\cite{mb1995}, modeled by an Ising ferromagnet ($J<0$ in Eq.~(\ref{eqn1})). The MB algorithm accelerates growth by suppressing diffusion along the interfaces of differently ordered domains and favoring intra-domain bulk diffusion. This is because, in the late stages of evolution, only a small fraction of time is spent on intra-domain transport processes, which result in dynamical scaling and asymptotic growth law. Here, we modify the MB algorithm for the order-disorder transition. In the modified approach, we keep track of the antiparallel coordination number of each site $i$,
\begin{eqnarray}
 \label{eqn4}
 Q(i) = \sum_{L_i}\delta(S_i,-S_{L_i}).
\end{eqnarray}
The $Q(i)$'s run from $0$ to $z$, where $z$ is the lattice coordination number. Here, $z=4$ for $d=2$ square lattices. In Eq.~(\ref{eqn4}), $L_i$ denotes the neighbors of $i$. These $Q(i)$'s provide sufficient information to compute energy changes due to spin exchanges. The change in energy resulting from the exchange of two nn spins $i$ and $j$ of opposite sign is given by
\begin{eqnarray}
 \label{eqn5}
 \Delta E = 4J\left[ Q(i) + Q(j) - (z+1)\right].
\end{eqnarray}

We order all the sites having equal $Q(i)$ into lists. Thus, we have $z+1$ lists and all the sites in a given list have an identical environment. When the system is quenched below $T_c$, ordering starts throughout the system. As a result, the size of the list with $Q(i)=z$ will increase by shrinking the size of the lists with $Q(i)<z$. The steps of the algorithm are the same as those proposed by MB~\cite{mb1995}. However, for the sake of completeness, we will mention the steps. We choose a step from the ensemble of all possible spin exchanges according to how likely it is to occur per unit time, making time steps of appropriate duration. One step of our dynamics for a cubic lattice consists of the following sequence:\\
1. We increment time by
 \begin{eqnarray}
  \label{eqn6}
  \Delta t = \left[ \sum_{q=0}^z \left( 1 - \frac{q}{z} \right) N_q e^{-4\beta Jq}\right]^{-1},
 \end{eqnarray}
 where $N_q$ is the number of elements in the list with $q$ opposite neighbors. We refer to this time as \textit{Monte Carlo time}~(MCT), which is distinct from MCS.\\
2. We select the list of $q$ opposite neighbors with the probability
 \begin{eqnarray}
  \label{eqn7}
  P_q = \Delta t\left( 1 - \frac{q}{z}\right) N_q e^{-4\beta Jq}.
 \end{eqnarray}
3. We randomly select a site $i$ from the list of $q$ opposite neighbors.\\
4. We randomly select a neighbor $j$ of site $i$ with $S_i\ne S_j$.\\
5. We exchange the spins $S_i$ and $S_j$ according to the probability given by Eq.~(\ref{mcs}). Then, adjust $Q$-values of the sites $i$, $j$ and their neighbors, and update the lists.

This algorithm is much faster than the standard MC algorithm as described earlier. Also, it yields a time evolution equivalent to the standard MC algorithm, but time is updated in non-uniform increments as given by Eq.~(\ref{eqn6}). Therefore, the MB algorithm enables us to access the asymptotic growth regime. We use this algorithm to obtain statistical data in the asymptotic regime. However, we only show the length-scale data in MCT, the ``time" in the MB algorithm.

\subsection{Detailed Results from MC Simulations}\label{3s2sec2}
\begin{figure}
\centering
\includegraphics*[width=0.60\textwidth]{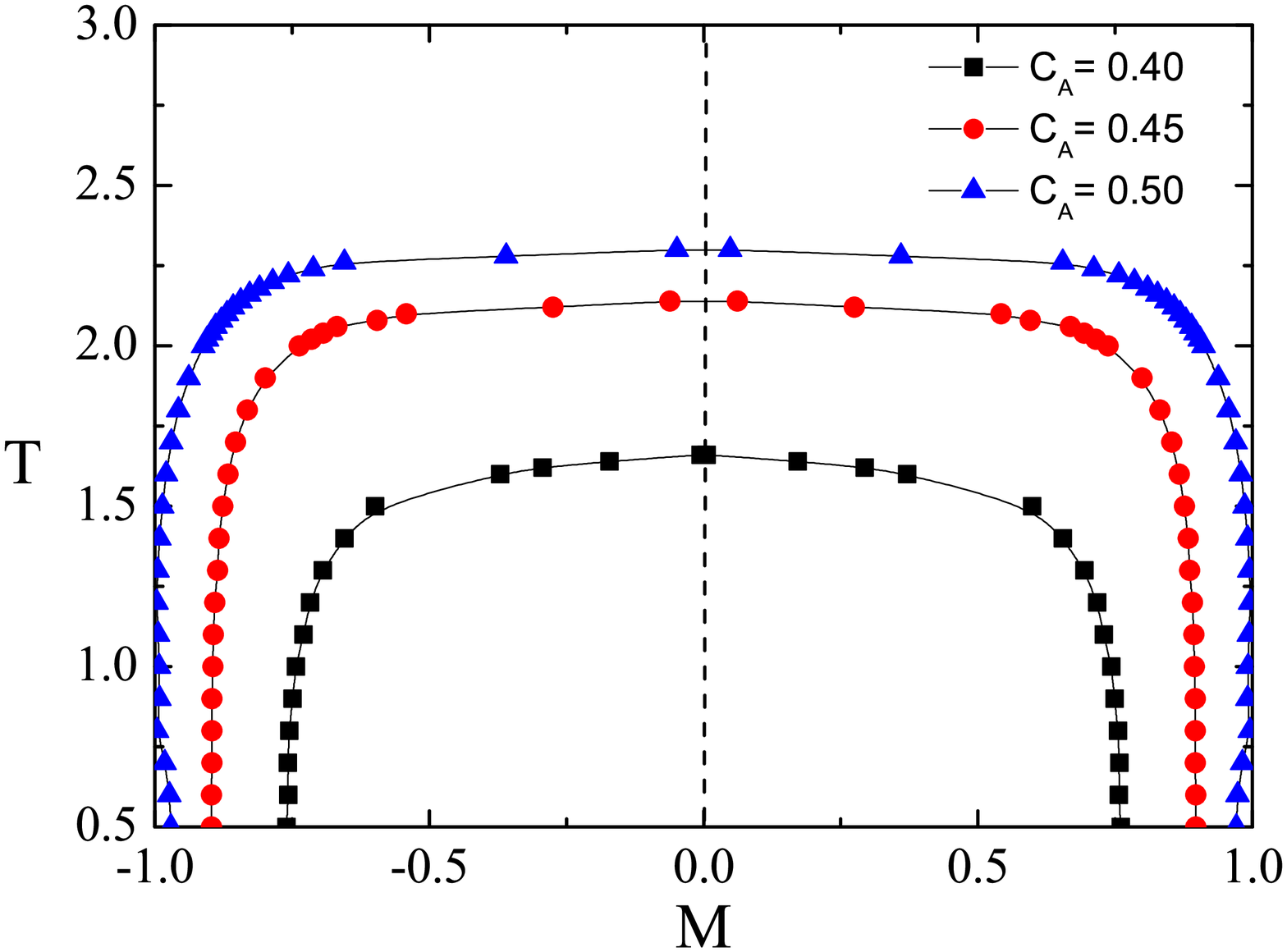}
\caption{\label{3fig1} Coexisting phases of order-disorder transition in symmetric and asymmetric binary mixtures below the critical temperature $T_c(c_A)$. We considered an ensemble of small Ising systems with fixed concentrations of A and B atoms and studied the equilibrium phases at different $T$. We set parameters $J$ and $k_B$ to unity. Clearly, as we increase the asymmetry in composition, i.e., by reducing $c_A$ from $0.50$, the transition temperature, $T_c(c_A)$, and the amplitude of the \textit{staggered magnetization}, $M$ in the bulk of ordered domains decreases from the symmetric mixture.}
\end{figure}

We obtain the equilibrium values of staggered magnetization at different $T$ by equilibrating relatively small systems. Figure~\ref{3fig1} shows the coexisting phases in ($M$, $T$) plane for various values of $c_A$. Details are given in the figure caption. The mean-field $T_c$ for the order-disorder transition is given by~\cite{mpbb06}
\begin{eqnarray}
\label{tc}
T_c(c_A) = \frac{4Jzc_A(1 - c_A)}{k_B}
\end{eqnarray}
We estimate numerical values of $T_c(c_A)$ from Fig.~\ref{3fig1} as the point at which $M$ drops to zero. Table~\ref{3table:1} shows the approximate value of $T_c(c_A)$ for different values of $c_A$.

\begin{table}
	\centering
	\begin{tabular}{| p{2.5cm} | p{3.0cm} |} 
		\hline  
		\hskip 1.0 cm $c_A$ & \hskip 1.2 cm $T_c$($J/k_B$) \\ [1ex] \hline
		\hskip 0.8 cm 0.40  & \hskip 0.8 cm 1.66332  \\ [1ex]  \hline
		\hskip 0.8 cm 0.45  & \hskip 0.8 cm 2.14293  \\ [1ex]  \hline
		\hskip 0.8 cm 0.50  & \hskip 0.8 cm 2.26931  \\ [1ex]  \hline
	\end{tabular}
	\caption{Critical temperature for the antiferromagnet at different values of $c_A$.}
	\label{3table:1}
\end{table}

\begin{figure}
\centering
\includegraphics*[width=0.65\textwidth]{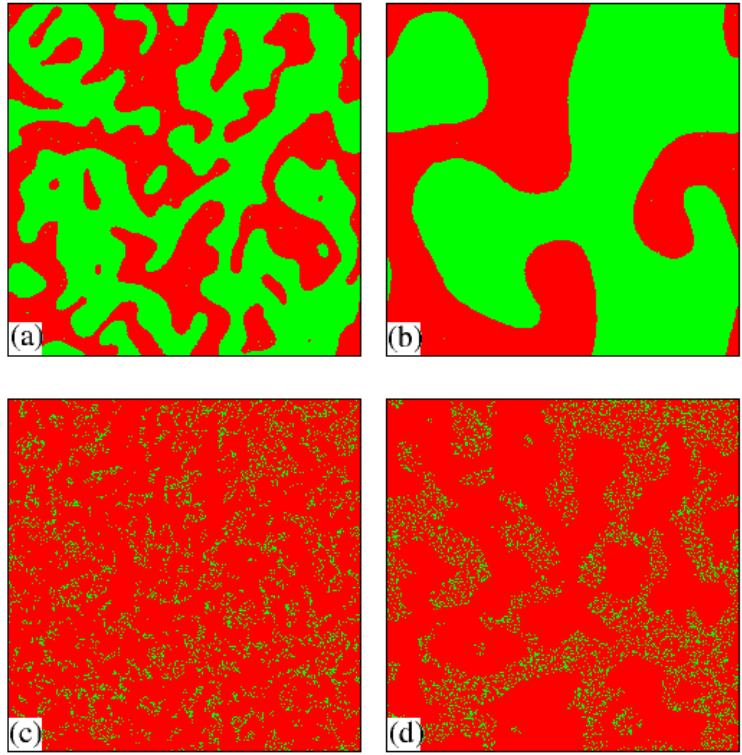}
\caption{\label{3fig2} Evolution snapshots of staggered magnetization $\sigma_{ij}$, obtained from the regular MC simulation of the Ising model in two dimensions. The system size is $4096^2$. First row shows the snapshots for mixtures with $c_A=0.50$ and second row shows the snapshots for $c_A=0.40$ at two different MCSs: $MCS=10^4$ for (a) and (c); and  $MCS=10^5$ for (b) and (d). The initial condition for each run consists of homogeneous mixture of A and B according to the desired ratio. The quenching temperature is $T=0.9$ for both the cases. Sites with $\sigma_{ij}>0$ are marked in red while sites with $\sigma_{ij}<0$ are marked in green. Evolution snapshots in the second row show that the excess B atoms were dissolved into the bulk.}
\end{figure}

In Fig.~\ref{3fig2}, we show the evolution snapshots of the staggered magnetization field $\sigma_{ij}=(-1)^{i+j}S_{ij}$ for $c_A=0.50$ and $c_A=0.40$ at different MCS, as specified. For the symmetric mixture with $c_A=c_B=0.50$, immediately after the quench below $T_c$, the system evolves toward two degenerate ABAB and BABA states. In the late stages, domain growth is driven by the removal of the interfaces between these two phases. But for the asymmetric mixture with $c_A=0.40$, ordering starts throughout the system and the excess B atoms start accumulating along the interfaces (wetting the interfaces) of the ordered regions. This wetting reduces the surface tension, slowing down domain growth at intermediate times, which could be misinterpreted as a lower exponent than $1/2$ or a logarithmic growth. However, in the late stages of domain growth, excess B atoms start migrating into the bulk of the ordered domains. The bulk domains then settle to their equilibrium composition with the ABAB/BABA structure interspaced with surplus B-atoms, reducing the staggered magnetization (Fig.~\ref{3fig1}). This scenario is clear from the evolution snapshots for $c_A=0.40$ in Fig.~\ref{3fig2}. For all other asymmetric compositions, similar dynamics is observed, except the intermediate wetting regime becomes more prolonged for greater asymmetry.

In order to study the morphology of domain growth, we calculate the correlation function and structure factor of the staggered magnetization field, $\sigma_{ij}$. The equal-time correlation function $C(\vec r, t)$ is defined as follows
\begin{eqnarray}
\label{eqn2}
C\left(\vec r, t\right) = \left[\left\langle \sigma(\vec R, t)\sigma(\vec R + \vec r, t) \right\rangle -  \left\langle\sigma(\vec R, t)\right\rangle\left\langle\sigma(\vec R + \vec r, t) \right\rangle\right].
\end{eqnarray}
Here, the angular brackets represent an average over different initial conditions~\cite{pw2009}. Similarly, we calculate the {\it structure factor} $S(k, t)$, which is defined as the Fourier transform of $C(r, t)$:
\begin{eqnarray}
\label{eqn3}
S\left(\vec k, t\right) = \int d\vec r e^{i\vec k.\vec r}C(\vec r, t)
\end{eqnarray}
at wave vector $\vec k$. For ordering in the 50\%-50\% mixture, we expect the correlation function to obey the OJK theory \cite{ojk82}. The OJK theory studies the nonconserved ordering of a ferromagnet via defect dynamics~\cite{pw2009}. The functional form of the OJK function is 
\begin{eqnarray}
\label{ojk}
C_\text{OJK}\left(r, t\right) = \frac{2}{\pi}\sin^{-1}\left(e^{-r^2/L_\sigma^2}\right), ~~~~~~~L_\sigma\sim t^{1/2} .
\end{eqnarray}

\begin{figure}
\centering
\includegraphics*[width=0.70\textwidth]{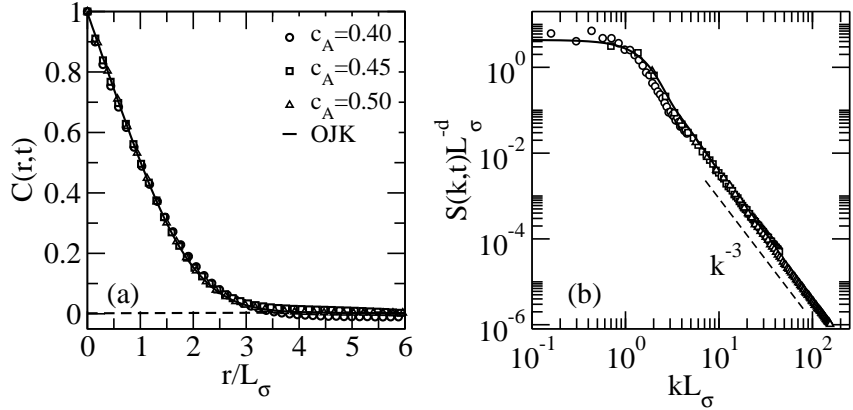}
\caption{\label{3fig3} Scaling plot of spherically averaged correlation functions $C(r, t)$ and structure factors $S(k, t)$ of $\sigma(\vec r, t)$ field at $10^5$ MCS. Results are obtained by averaging over ten independent runs. (a) Plot of $C(r, t)$ vs. $r/L_\sigma$ for different $c_A$, as mentioned. We define the length scale, $L_\sigma$, as the distance at which $C(r, t)$ falls to half from its maximum ($=1$ at $r=0.0$). The solid line corresponds to the OJK function in Eq.~(\ref{ojk}), collapsed appropriately with the numerical data. (b) Plot of $S(k, t)L_\sigma^{-d}$ vs. $kL_\sigma$ on a {\it log-log} scale. The solid line represents the Fourier transform of the OJK function and it is indistinguishable from the numerical data. The dashed line labeled with $k^{-3}$ represents the Porod's law: $S(k,t)\sim k^{-(d+n)}$ with $d=2$ and $n=1$. The system size is $2048^2$ and rest of the details are same as given in the caption of Fig.~\ref{3fig2}.}
\end{figure}

In Fig.~\ref{3fig3}(a), we plot $C(r, t)$ vs. $r/L_\sigma$ at $10^5$ MCS for different values of $c_A$, as mentioned. We define $L_\sigma$ as the distance over which $C(r, t)$ falls to $1/2$ of its maximum value [$C(0, t)=1$]. We also plot the OJK function in Eq.~(\ref{ojk}). We see that numerical data for all values of $c_A$ are indistinguishable from the OJK function, confirming that the asymptotic morphology of ordering in symmetric and asymmetric binary mixtures is same. The only difference between the symmetric and asymmetric cases is the purity of the ordered domains, i.e., the value of $\langle\sigma\rangle$.

In Fig.~\ref{3fig3}(b), we plot $S(k, t)L_\sigma^{-d}$ vs. $kL_\sigma$ at $10^5$ MCS for different values of c$_A$, as specified. Again, the scaling functions are indistinguishable from each other. We also plot the Fourier transform of the OJK function, which is in excellent agreement with our numerical data. In the limit $k\rightarrow\infty$, $S(k, t)$ decays as $k^{-3}$, following the {\it Porod's law}~\cite{gp1982,yosp88}. This results from scattering off sharp interfaces formed between the two degenerate ordered states, irrespective of the amount of surplus material present in the system. Notice that, if the surplus material wets the interfaces, they are no longer sharp - this would interfere with the observation of the Porod tail~\cite{gp1982,yosp88}.

\begin{figure}
\centering
\includegraphics[width=0.45\textwidth]{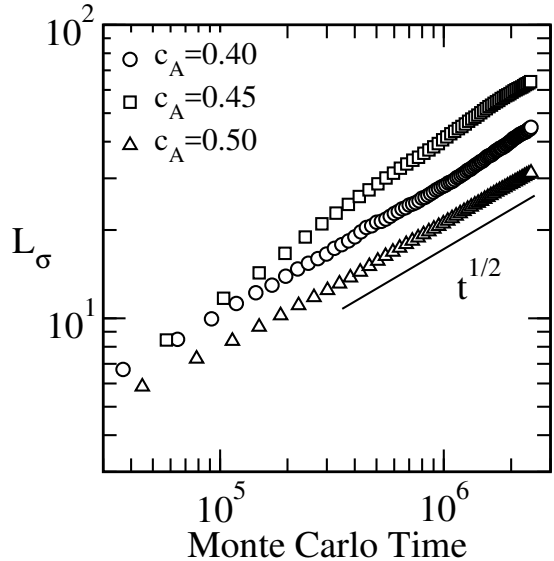}
\caption{\label{3fig4} Time-dependence of the characteristic length scales $L_\sigma$ of $\sigma(\vec r, t)$ field. Plot of $L_\sigma$ vs. $Monte~Carlo~time$ for different values of $c_A$, as specified. Results are obtained from the accelerated MC simulation in two dimensions. The system size is $2048^2$ and quench temperature is $T=0.9$ for all the mixtures. The solid line labeled with $t^{1/2}$ represents the Allen-Cahn growth law. Clearly, the growth exponent is same for symmetric and asymmetric in mixtures.}
\end{figure}

Finally, let us present data for the time-dependence of the length scale $L_\sigma$. To unambiguously access the asymptotic regime, we studied the ordering by using the accelerated MC algorithm, as described in Sec.~\ref{3s1sec2}. The variation of $L_\sigma$ vs. MCT for different $c_A$'s is shown in Fig.~\ref{3fig4}. The solid line with exponent $\alpha=1/2$ corresponds to diffusive growth for nonconserved order parameter, and denotes the well-known {\it Allen-Cahn}~(AC) Law: $L_\sigma(t) \sim t^{1/2}$~\cite{ca1979}. For the symmetric mixture with $c_A=0.50$, the growth law is consistent with the AC law from early times. For asymmetric mixtures, there is an initial regime of slower growth due to the reduction in surface tension by surface wetting. Even at later times, the purity of bulk domains diminishes due to surplus B-atoms. This reduces the average magnetization, and the surface tension between coexisting phase. However, the asymptotic behavior always shows a crossover to the AC regime, regardless of the composition of the mixture. All information about the composition is contained in the prefactor of the growth law.

From the evolution snapshots shown in Fig.~\ref{3fig2}, it is clear that the average length-scale is larger for the symmetric mixture at a given MCS. However, the length-scale data shown in Fig.~\ref{3fig4} is ordered differently. This is because the proportionality constant between MCT and MCS is different for different mixtures, and it is largest for the symmetric mixture. These constants affect the prefactor of the growth law.

\section{Coarse-Grained Model and Numerical Results}\label{3sec3}
\subsection{Details of Coarse-Grained Model}\label{3s1sec3}
We obtain an approximate coarse-grained free energy functional $\mathcal{F}\left[m, \psi\right]$ to describe the ordering kinetics in binary alloys from the Hamiltonian. It consists of two order parameters. One is the coarse-grained staggered magnetization, $m(\vec r, t)$ which is a nonconserved order parameter, and the other is the local concentration difference of the mixture compositions, $\psi(\vec r, t)$ which is a conserved order parameter. The time evolution of these order-parameter fields are described by the \textit{Model C} in Hohenberg and Halperin's nomenclature~\cite{HH1977}.

Consider a binary mixture (A+B) in a simple cubic lattice. Let $N$ be the total number of lattice sites and $z$ is the number of nn sites around each site. $N_A$ and $N_B$ are the number of A and B atoms in the mixture. Next, we divide the lattice into two sublattices where $N_{A1}$ ($N_{A2}$) and $N_{B1}$ ($N_{B2}$) are, respectively, the number of A and B atoms in the sublattice 1(2). Then, we have the following relations
\begin{align}
 \label{3cgeq2}
 N_{A1} + N_{A2} &= N_A = c_AN,\nonumber\\
 N_{B1} + N_{B2} &= N_B = c_BN,\nonumber\\ 
 N_{A1} + N_{B1} &= N_{A2} + N_{B2} = N/2,\nonumber\\
 c_A + c_B &= 1.
\end{align}
For the sake of simplicity, we consider $N_A\leq N_B$, and define the nonconserved order parameter as
\begin{eqnarray}
 \label{3cgeq3}
 m = \frac{N_{A1} - N_{A2}}{N_A}.
\end{eqnarray}
Thus, we have $-1\leq m \leq 1$. We also define the conserved order parameter as
\begin{eqnarray}
 \label{3cgeq4}
 \psi = c_A - c_B.
\end{eqnarray}
Therefore, the regions with $\psi=+1$ (or $-1$) correspond to A-rich (or B-rich) domains. Using Eqs.~(\ref{3cgeq2}), (\ref{3cgeq3}) and (\ref{3cgeq4}), the sublattice occupancies are obtained as follows
\begin{align}
 \label{3cgeq5}
 N_{A1} &= \frac{1}{2}(1 + m)N_A = \frac{N}{4}(1 + \psi)(1 + m),\nonumber\\
 N_{A2} &= \frac{1}{2}(1 - m)N_A = \frac{N}{4}(1 + \psi)(1 - m),\nonumber\\
 N_{B1} &= \frac{1}{2}(N_B - mN_A) = \frac{N}{4}[(1 - \psi) - m(1 + \psi)],\nonumber\\
 N_{B2} &= \frac{1}{2}(N_B + mN_A) = \frac{N}{4}[(1 - \psi) + m(1 + \psi)].
\end{align}

If $N_{AA}$, $N_{BB}$ and $N_{AB}$ are, respectively, the number of AA, BB and AB  (or BA) type of bonds in the system, then the interaction energy of this configuration is
\begin{eqnarray}
 \label{3cgeq1}
 E = N_{AA}e_{AA} + N_{BB}e_{BB} + N_{AB}e_{AB},
\end{eqnarray}
where $e_{AA}$, $e_{BB}$ and $e_{AB}$ are, respectively, the strength of AA, BB and AB(or BA) type of bonds. In terms of sublattice occupancies, the approximate number of bonds are given by
\begin{align}
 \label{3cgeq6}
 N_{AA} &= \frac{zN_{A1}N_{A2}}{N/2} = \frac{zN}{8}(1 + \psi)^2(1 - m^2),\nonumber\\
 N_{BB} &= \frac{zN_{B1}N_{B2}}{N/2} = \frac{zN}{8}\left[(1 - \psi)^2 - m^2(1 + \psi)^2)\right],\nonumber\\
 N_{AB} &= \frac{z}{N/2}(N_{A1}N_{B2} + N_{B1}N_{A2}) = \frac{zN}{4}\left[(1 - \psi^2) + m^2(1 + \psi)^2\right].
\end{align}
Using Eqs.~(\ref{3cgeq6}), the expression of $E$ in Eq.~(\ref{3cgeq1}) reduces to
\begin{eqnarray}
 \label{3cgeq7}
 E = \frac{zN\epsilon}{4}(\psi^2 - m^2 -2m^2\psi -m^2\psi^2)+\frac{zN}{4}(e_{AA}-e_{BB})\psi+\frac{zN}{8}(e_{AA}+e_{BB}+2e_{AB}),
\end{eqnarray}
where $\epsilon = \frac{1}{2}(e_{AA} + e_{BB}) - e_{AB}$. Next, the entropy of the configuration is given by
\begin{eqnarray}
 \label{3cgeq8}
 S=-k_B\left(N_{A1}\ln\frac{N_{A1}}{N/2} + N_{A2}\ln\frac{N_{A2}}{N/2} + N_{B1}\ln\frac{N_{B1}}{N/2} + N_{B2}\ln\frac{N_{B2}}{N/2}\right).
\end{eqnarray}

The Bragg-Williams free energy of the system can be written as (dropping constant terms and using Eq.~(\ref{3cgeq5}))
\begin{align}
\label{3cgeq9}
\nonumber
\frac{F}{N} =& \frac{E - TS}{N}\nonumber \\
  =& \frac{z\epsilon}{4}(\psi^2 - m^2 -2m^2\psi -m^2\psi^2) \nonumber \\
   & + \frac{k_BT}{2}\left\{\frac{(1+\psi)(1+m)}{2}\ln\frac{(1+\psi)(1+m)}{2} + \frac{(1+\psi)(1-m)}{2}\ln\frac{(1+\psi)(1-m)}{2} \right.\nonumber \\ & \left. + \frac{(1-\psi)-m(1+\psi)}{2}\ln\frac{(1-\psi)-m(1+\psi)}{2}  +\frac{(1-\psi)+m(1+\psi)}{2}\ln\frac{(1-\psi)+m(1+\psi)}{2}\right\}.
\end{align}
Expanding the logarithmic terms in Eq.~(\ref{3cgeq9}) up to fourth order in $m$ and $\psi$, we obtain the following form of approximate free energy
\begin{eqnarray}
 \label{3cgeq10}
 \frac{F}{N} &=& \left(\frac{z\epsilon}{4} + \frac{k_BT}{2}\right)\psi^2 + \left(\frac{k_BT}{2} - \frac{z\epsilon}{4}\right)m^2 + \frac{k_BT}{12}\psi^4\nonumber\\
  & & + \frac{k_BT}{12}m^4 + \left(k_BT - \frac{z\epsilon}{2}\right)\psi m^2 + \left(k_BT - \frac{z\epsilon}{4}\right)\psi^2m^2.
\end{eqnarray}
Let $a = \frac{z\epsilon}{2} + k_BT$, $b = \frac{z\epsilon}{2} - k_BT$, $c = \frac{k_BT}{3}$, and $d = 2k_BT - \frac{z\epsilon}{2}$. Therefore, from Eq.~(\ref{3cgeq10}), the expression of free energy density can be written as
\begin{eqnarray}
 \label{3cgeq14}
 f\left(m, \psi\right) = \frac{a}{2}\psi^2 - \frac{b}{2}m^2 + \frac{c}{4}m^4 - b\psi m^2 + \frac{d}{2}\psi^2m^2.
\end{eqnarray}
Again, since $a$, $b$, $c$, and $d$ are all positive, we dropped the $\psi^4$ term in Eq.~(\ref{3cgeq14}).

We consider the following form of the GL free energy functional 
$\mathcal{F}\left[m, \psi\right]$:
\begin{eqnarray}
 \label{3cgeq16}
 \mathcal{F}\left[m, \psi\right] = \int d\vec r\left[ f\left(m, \psi\right) + \frac{K}{2}\left(\vec\nabla m\right)^2 \right],
\end{eqnarray}
where $K$ is the energy cost due to the spatial variation of $m(\vec r, t)$. In the microscopic theory, parameters $a$, $b$, $c$, and $d$ are dependent on each other. However, in the GL formalism, we will treat them as independent. In Eq.~(\ref{3cgeq14}), the second and third terms correspond to a double-well potential for $m(\vec r, t)$ field. The locations of minima depend on the mean value of $\psi(\vec r, t)$. The presence of the coupled term $m^2\psi^2$ effectively reduces the critical temperature of the mixture as the mixture becomes more and more asymmetric in composition. Therefore, the free energy in Eq.~(\ref{3cgeq14}) contains all the features as observed in the microscopic dynamics. 

Here, we described the time evolution of the order-parameters~\cite{pw2009,sdsp04,ajb94,HH1977}. Since $m(\vec r, t)$ is a nonconserved quantity, its time evolution is given by the \textit{time-dependent Ginzburg-Landau} (TDGL) equation as 
\begin{eqnarray}
 \label{3eqn10}
 \frac{\partial m}{\partial t} &=& -\varGamma_m\left(\frac{\delta \mathcal{F}}{\delta m}\right) + \eta_m, \nonumber \\
 &=&-\varGamma_m\left( -bm + cm^3 + dm\psi^2 - 2bm\psi - K\nabla^2 m\right) + \eta_m,
\end{eqnarray}
where $\varGamma_m$ is the kinetic coefficient and $\eta_m$ represents the thermal noise. Again, the time evolution of the conserved $\psi(\vec r, t)$ field is described by the \textit{Cahn-Hilliard-Cook}(CHC) equation as
\begin{eqnarray}
 \label{3eqn11}
 \frac{\partial\psi}{\partial t} &=& \varGamma_\psi\nabla^2\left(\frac{\delta \mathcal{F}}{\delta \psi}\right) + \eta_\psi,  \nonumber \\
 &=& \varGamma_\psi\nabla^2\left( a\psi - bm^2 + d\psi m^2 \right) + \eta_\psi,
\end{eqnarray}
where $\varGamma_\psi$ is the kinetic coefficient and $\eta_\psi$ represents the thermal noise.

To obtain the dimensionless version of Eqs.~(\ref{3eqn10}) - (\ref{3eqn11}), we consider following transformations
\begin{eqnarray}
 \label{3trans}
 m = m_0m', \hskip 0.4cm \psi = \psi_0\psi', \hskip 0.4cm t = t_0t', \hskip 0.4cm \vec r = r_0\vec r^\prime,
\end{eqnarray}
where all the dimensionless variables carry prime sign. Let,
\begin{eqnarray}
 \label{3vari}
 m_0 = \sqrt{\frac{b}{c}}, \hskip 0.4cm \psi_0 = \frac{b^2}{ac}, \hskip 0.4cm t_0 = \frac{1}{b\varGamma_m}, \hskip 0.4cm r_0 = \sqrt{\frac{K}{b}}.
\end{eqnarray}
Inserting Eqs.~(\ref{3trans}) - (\ref{3vari}) into Eqs.~(\ref{3eqn10}) - (\ref{3eqn11}), we obtain the dimensionless (dropping prime signs and introducing $\varGamma = \frac{a\varGamma_\psi}{b\varGamma_m}$, $\alpha = \frac{b^2}{ac}$ and $\beta = \frac{bd}{ac}$) version of the evolution equations as
\begin{eqnarray}
 \label{3final_m}
 \frac{\partial m}{\partial t} &=& m - m^3 + 2\alpha m\psi -\alpha\beta m\psi^2 + \nabla^2m,\\
 \label{3final_p}
 \frac{\partial \psi}{\partial t} &=& \varGamma\nabla^2\left[\psi - m^2 + \beta\psi m^2\right].
\end{eqnarray}
These two nonlinear coupled evolution equations for $m(\vec r, t)$ field and $\psi(\vec r, t)$ field describe the ordering kinetics in binary mixtures.

\subsection{Results from Coarse-Grained Simulations}\label{3s2sec3}
Here, we discuss numerical results obtained from the coarse-grained model. We numerically solve Eqs.~(\ref{3final_m}) - (\ref{3final_p}) by using Euler discretization with spatial mesh sizes $\Delta x=1.0$ and time step $\Delta t=0.01$ respectively. The system size is $L_x\times L_y=4096\times 4096$. We employ periodic boundary conditions in all the directions for both $m(\vec r, t)$ and $\psi(\vec r, t)$, respectively. The initial condition for $m(\vec r, t)$ field is $m(\vec r, 0) = \pm 0.01$ which corresponds to the disordered state before the quench. Similarly, the initial condition for $\psi(\vec r, t)$ field is $\psi(\vec r, 0) = \psi_m \pm 0.01$, where $\psi_m$ is the mean concentration of surplus material in the system. We choose numerical values of the coefficients as $\varGamma = 1$, $\alpha = 1$, and $\beta = 1$ respectively.

\begin{figure}
\centering
\includegraphics*[width=0.8\textwidth]{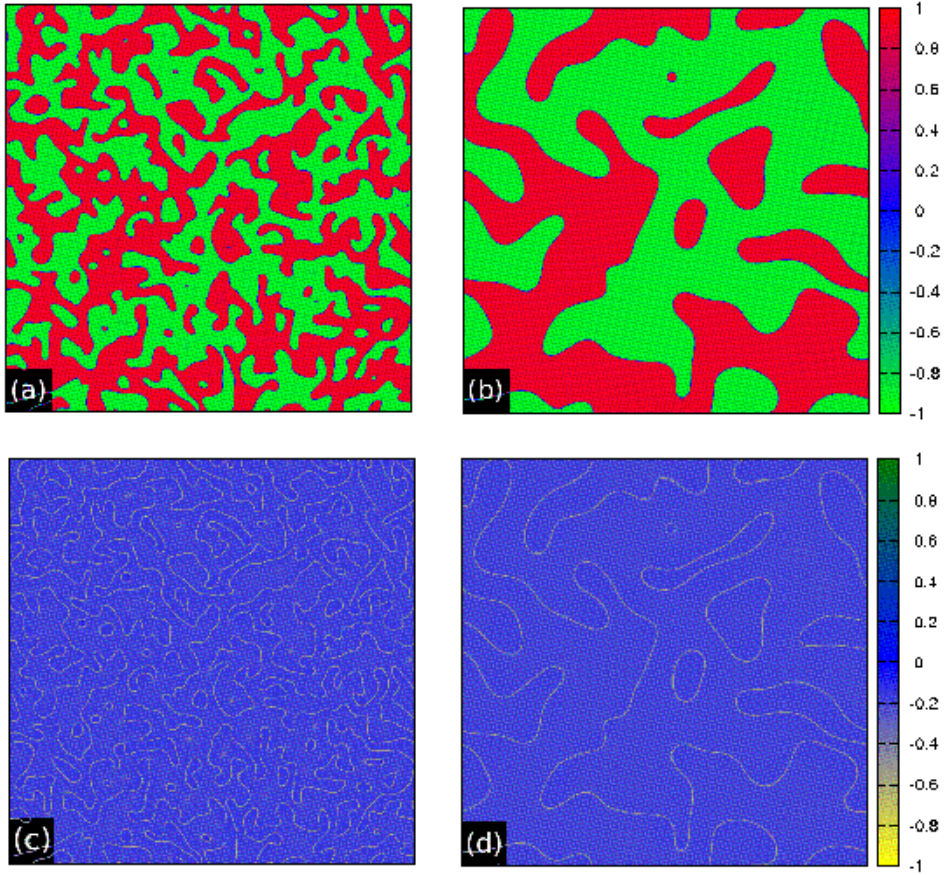}
\caption{\label{3fig6} Evolution snapshots of the nonconserved $m(\vec r, t)$ field and the conserved $\psi(\vec r, t)$ field in $d=2$: (a) $m(\vec r, t)$ field at $t=1000$, (b) $m(\vec r, t)$ field at $t=10000$, (c) $\psi(\vec r, t)$ field at $t=1000$, and (d) $\psi(\vec r, t)$ field at $t=10000$. Results are obtained from coarse-grained simulations described by Eqs.~(\ref{3final_m}) and (\ref{3final_p}) respectively for $\psi_m = -0.20$. The lattice size was $L^2 = 4096^2$. For the sake of clarity, we have shown only $2048^2$ corner of the simulation area. Periodic boundary conditions were applied in all directions. Color bars in the first and second rows are, respectively, represent the amplitude of $m(\vec r, t)$ and $\psi(\vec r, t)$ fields. Clearly, except at the boundaries of two different phases of $m(\vec r, t)$ field, $\psi(\vec r, t)$ field is uniform through out the system.}
\end{figure}

Figure~\ref{3fig6} shows the evolution snapshots of the order parameter fields for $\psi_m = -0.20$ (equivalent to the asymmetric mixture with $c_A = 0.4$ and $c_B = 0.6$ in the MC simulation) at different times, as mentioned. In Figs.~\ref{3fig6}(a) and \ref{3fig6}(b), we plot the time evolution of $m(\vec r, t)$ field at different times (details are given in the figure caption). Clearly, as time advances, the average domain size increases by removing the interfaces between the two degenerate phases. The color bar adjacent to Fig.~\ref{3fig6}(b) shows the amplitude of the $m(\vec r, t)$ field at different space points which is less than unity. Figures~\ref{3fig6}(c) and \ref{3fig6}(d) represent the time evolution of the $\psi(\vec r, t)$ field. The adjacent color bar to Figs.~\ref{3fig6}(d) shows that $\psi(\vec r, t)$ is uniform throughout the system with $\psi(\vec r, t)\approx\psi_m$, apart from the regions where the domain walls of the $m(\vec r, t)$ field are located. This implies that the domain wall between the two degenerate phases of the $m(\vec r, t)$ field is sharp, and it is similar to what observed in regular MC simulation.

\begin{figure}
\centering
\includegraphics*[width=0.60\textwidth]{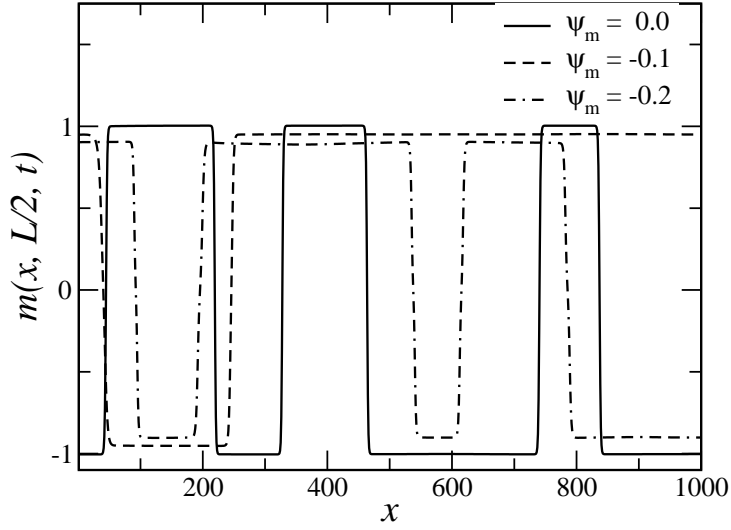}
\caption{\label{3fig7} The profile of staggered magnetization $m(\vec r, t)$ field along the $y = L/2$ line of the simulation box at $t=5000$ for different values $\psi_m$, as specified. We have shown data up to $x=1000$. Clearly, the amplitude of $m(\vec r, t)$ field decreases with the increase $|\psi_m|$, i.e., the increase of asymmetry in composition. Rest of the simulation details are same as given in the caption of Fig.~\ref{3fig6}.}
\end{figure}

In Fig.~\ref{3fig7}, we plot the profile of the $m(\vec r, t)$ field along the $y=L/2$ line of the simulation box at $t=5000$ for different values of $\psi_m$, as specified. For $\psi_m = 0$, which corresponds to the symmetric mixture with $c_A = c_B = 0.5$ in the MC simulation, the amplitude of $m(\vec r, t)$ field in the bulk of ordered domains is $m_b=\pm1$. For all other cases with $\psi_m<0$, which correspond to asymmetric mixtures with $c_A < c_B$ in the MC simulation, $m_b$ is less than unity. Clearly, higher the asymmetry smaller is the $|m_b|$. The coexisting phases obtained from the MC simulation in Fig.~\ref{3fig1} qualitatively supports these results. 

\begin{figure}
\centering
\includegraphics*[width=0.60\textwidth]{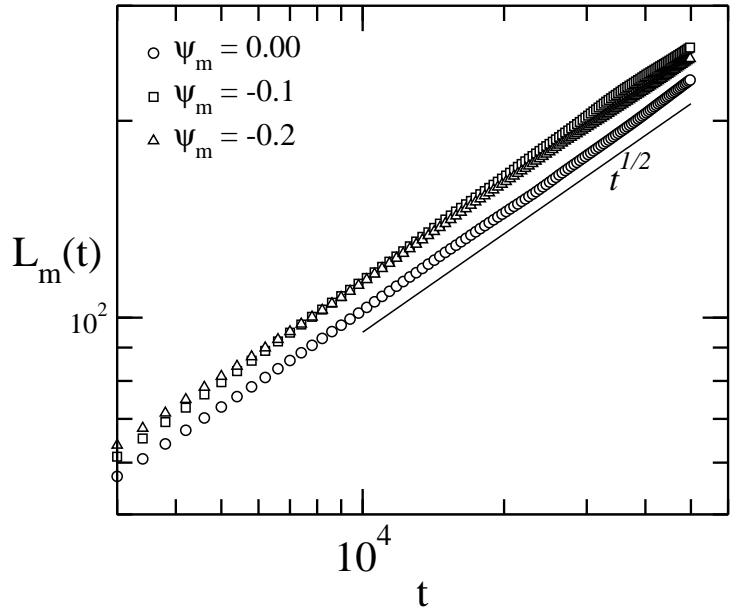}
\caption{\label{3fig8} Time-dependence of the characteristic length scale $L_m(t)$ of $m(\vec r, t)$ field for different values of $\psi_m$, as mentioned. Plot of $L_m(t)$ vs. $t$ on {\it log-log} scale. Line labeled with $t^{1/2}$ corresponds to diffusive growth. Clearly, the growth exponent is same for all values of $\psi_m$.}
\end{figure}

Finally, we calculate the average domain size $L_m(t)$ of $m(\vec r, t)$ field from the equal-time correlation function,
\begin{eqnarray}
\label{3cormm}
C_{mm}\left(\vec r, t\right) = \left[\left\langle m(\vec R, t)m(\vec R + \vec r, t) \right\rangle -  \left\langle m(\vec R, t)\right\rangle\left\langle m(\vec R + \vec r, t) \right\rangle\right].
\end{eqnarray}
We define the distance $r=L_m$ at which $C_{mm}\left(r, t\right)$ falls to half of its maxima [$1$ at $r=0$]. In Fig.~\ref{3fig8}, we plot $L_m(t)$ vs. $t$ for different values of $\psi_m$, as mentioned. The solid line labeled with $t^{1/2}$ corresponds to the AC law. For $\psi_m = 0$, we obtain $L_m(t)\sim t^{1/2}$ throughout the simulation time. For all the cases with $\psi_m<0$, we find $L_m(t)\sim t^{1/2}$ in the late stages. At early times, we observe slower growth with $\alpha<1/2$. This is due to the wetting of domain boundaries of $m(\vec r, t)$ field by the excess B atoms, which effectively reduces the surface tension. These results confirm that the asymptotic domain growth law for ordering in symmetric and asymmetric mixtures follow the universal growth law: $L_m(t)\sim t^{1/2}$.

\section{Conclusion and Outlook}\label{3sec4}
Let us conclude this paper with a summary and discussion of our results. We have studied phase ordering kinetics in symmetric and asymmetric binary mixtures (AB) with AB-exchange kinetics. Microscopically, the dynamics is studied by using the antiferromagnetic Ising model with Kawasaki spin-exchange kinetics. The equal-time correlation functions of the {\it staggered magnetization} of symmetric and asymmetric mixtures show data collapse, indicating the morphological similarity of domain growth. The structure factor tail decays as $S(k, t)\sim k^{-3}$ (where $k$ is the wave vector) for all compositions. This results from scattering off sharp interfaces formed between two degenerate phase of staggered magnetization field, irrespective of the composition asymmetry. The late-stage domain growth exponent is always $\alpha=1/2$ even when the composition deviates from $c_A=0.50$. This is confirmed by using an accelerated Monte Carlo algorithm. For an asymmetric mixture, the excess material initially wets the domain boundaries. This reduces surface tension and yields slow transient behavior. At later times, the surplus atoms dissolved into the bulk of differently ordered domains. This corresponds to the asymptotic regime which is universal across mixture compositions. 

We also obtained the coarse-grained {\it Ginzburg-Landau}~(GL) free-energy functional from the Hamiltonian. The GL functional depends upon two order-parameter fields: the staggered magnetization $m(\vec r, t)$, which has a nonconserved kinetics; and the composition $\psi(\vec r, t)$ which obeys a conserved kinetics. The TDGL and CHC equations, respectively, describe the time evolution of $m(\vec r, t)$ and $\psi(\vec r, t)$. The amplitude of $m(\vec r, t)$ in the bulk of ordered domains decreases with the increase of the asymmetry in composition. The length scale of the $m(\vec r, t)$ field grows as $L_m(t)\sim t^{1/2}$, regardless of the asymmetry in composition. These results confirm the universality in the ordering of symmetric and asymmetric binary mixtures.

While in this study, we have focused on a two-dimensional model system, in future we would like to extend our study to realistic systems in three dimensions, so that our obtained results can be directly compared with experiment. Specifically, we will employ the density functional theory~(DFT) based quantum-chemical calculations of a given alloy system for realistic modeling of the underlying Ising-type Hamiltonian~\cite{tsd}. In this respect, we will consider FCC lattice based alloys like Pt$_3$Cu~\cite{cupt}, Cu$_3$Au~\cite{cuau}, and Cu$_3$Pd~\cite{cupd} which have been reported to order in L1$_2$ structure. High resolution electron microscopy experiments on Cu-17$\%$Pd alloy suggests initial phase of wetting of the boundaries between ordered domains~\cite{anik} as found in the present study. Similarly, we would like to study the transition between the B2 and DO$_3$ phase of BCC lattice based Fe-Al system, for example Fe-27$\%$Al for which high resolution electron microscopy experimental data exists~\cite{anik1}. \\
\ \\
\noindent{\bf Acknowledgments:} PD acknowledges financial support from Council of Scientific and Industrial Research, India. SP is grateful to Department of Science and Technology, India for support via a J.C. Bose fellowship. \\
\ \\
\noindent{\bf Statement of Contribution:} SP and TSD proposed the problem. PD performed the analytical and numerical work with the help of SP and TSD. The paper was written by all three authors.

\newpage

\end{document}